\documentclass[useAMS,usenatbib]{mn2e}
\usepackage{rotating}
\usepackage{booktabs}
\usepackage{setspace}
\usepackage{txfonts}
\usepackage{float}
\usepackage{amssymb}
\usepackage{epsfig}
\usepackage{color}
\usepackage{graphicx}
\usepackage{multirow}
\usepackage{longtable}
\usepackage{verbatim}
\usepackage[flushleft]{threeparttable}
\usepackage{gensymb}
\usepackage{epstopdf}

\usepackage{array}
\usepackage{natbib}

\newcolumntype{L}[1]{>{\raggedright\let\newline\\\arraybackslash\hspace{0pt}}m{#1}}
\newcolumntype{C}[1]{>{\centering\let\newline\\\arraybackslash\hspace{0pt}}m{#1}}
\newcolumntype{R}[1]{>{\raggedleft\let\newline\\\arraybackslash\hspace{0pt}}m{#1}}

\newcommand{\Nu}{\it{NuSTAR}}

\title[The reflection spectrum of 4U~1636$-$53]
{
\begin{center}
The reflection spectrum of the low-mass X-ray binary 4U~1636$-$53
\end{center}
}

\author[Yanan Wang et al.]{Yanan Wang$^{1}$\thanks{E-mail: yanan@astro.rug.nl}, Mariano M\'endez$^{1}$, Andrea Sanna$^{2}$,
Diego Altamirano$^{3}$, T. M. Belloni$^{4}$\\
$^{1}$Kapteyn Astronomical Institute, University of Groningen, PO BOX 800, NL-9700 AV Groningen, the Netherlands\\
$^{2}$Dipartimento di Fisica, Universit\'{a} degli Studi di Cagliari, SP Monserrato-Sestu km 0.7, I-09042, Monserrato, Italy\\
$^{3}$Department of Physics and Astronomy, University of Southampton, Highfield SO17 IBJ, UK\\
$^{4}$Istituto Nazionale di Astrofisica, Osservatorio Astronomico di Brera, Via E. Bianchi 46, I-23807 Merate, Italy\\}

\begin{document}

\date{Accepted ? December ?. Received ? December ?; in original form ? December ?}

\maketitle

\label{firstpage}

\begin{abstract}
We present 3$-$79~keV {\Nu} observations of the neutron star low-mass X-ray binary 4U~1636$-$53 in the soft, transitional and hard state.
The spectra display a broad emission line at 5$-$10~keV.
We applied several models to fit this line: A {\sc gaussian} line,
a relativistically broadened emission line model, {\sc kyrline}, and two models including relativistically smeared and ionized reflection 
off the accretion disc with different coronal heights, {\sc relxill} and {\sc relxilllp}. 
All models fit the spectra well, however, the {\sc kyrline} and {\sc relxill} models yield an
inclination of the accretion disc of $\sim88\degree$ 
with respect to the line of sight, which is at odds with the fact that this source shows no dips or eclipses. 
The {\sc relxilllp} model, on the other hand, gives a reasonable inclination of $\sim56\degree$.
We discuss our results for these models in this source
and the possible primary source of the hard X-rays. 
\end{abstract}

\begin{keywords}
accretion, accretion disc--binaries: 
X-rays: individual (4U~1636$-$53)
\end{keywords}

\section{INTRODUCTION}
The advent of moderate/high resolution and high effective area X-ray instruments in the last decade has provided numerous examples of reflection spectra
in low-mass X-ray binaries (LMXBs; e.g. \citealt{Cackett2008}) and active galactic nuclei (AGN, e.g. \citealt{Parker2014}).
The X-ray reflection component produced at the inner edge of the ion 
disc in these systems is due to fluorescence and Compton scattering \citep[e.g.][]{Guilbert1988,Lightman1988,Fabian1989}.
The current paradigm is that a power-law component irradiates the surface of the accretion disc and
the X-ray photons then interact with the material producing diverse atomic features.
In the case of an accreting neutron star (NS), however, the emission from the NS surface/boundary
layer can as well irradiate the accretion disc \citep{Popham2001}.
Generally, the reflection spectrum contains a broad emission line in the 6.4$-$7.0~keV band due to iron,
plus a Compton back-scattering hump at $\sim$10$-$30~keV \citep[e.g.][]{Risaliti2013, Miller2013}.
Disc reflection spectra may provide a powerful probe of the ion geometry,
like the inner radius and inclination of the ion disc \citep{Fabian1989}.

4U~1636$-$53 is a NS low-mass X-ray binary classified as an atoll source \citep{Hasinger1989}
with an orbital period of $\sim$3.8~h \citep{Paradijs1990} and a companion star
of mass $\sim$0.4~$M_{\odot}$ \citep{Giles2002}, at a distance of 6~kpc \citep{Galloway2006}.
Besides the high variability \citep[e.g.][]{Altamorano2008, Sanna2014},
a broad and asymmetric emission line probably due to Fe-K has been observed in this system
\citep[e.g.][]{Pandel2008, Cackett2010}.
\cite{Pandel2008} reported relativistic lines in three {\it XMM-Newton} observations of 4U~1636$-$53 
and interpreted the line profile as due to the blending of at least two Fe-K lines from iron in different ionization states.
\cite{Cackett2010} analysed the spectra of 10 neutron star LMXBs, including 4U~1636$-$53, and found that the lines can be fitted 
equally well by a phenomenological and a reflection model in most cases. 
In their work, \cite{Cackett2010} employed a reflection model assuming illumination
by a blackbody component, implying the boundary layer illuminates a geometrically thin disc.

\cite{Ng2010} analysed the same spectra as \cite{Pandel2008} and \cite{Cackett2010},
but found that the lines could be fitted well with a {\sc gaussian} model, suggesting a symmetric line profile.
\cite{Ng2010} interpreted the line width as the result of broadening due to Compton 
scattering in the surface layers of the ion disc. 
The analyses of \cite{Pandel2008} and \cite{Ng2010} differ in some ways.
For instance, \cite{Ng2010} took pileup and background effects into account while \cite{Pandel2008} also fitted the 
simutaneous {\it Rossi X-ray Timing Explorer} ({\it RXTE}) observations and did not correct for pileup in their work,
which might also cause the difference of the line profiles because of the different continuum.

\cite{Sanna2013} analysed six {\it XMM-Newton} observations of 4U~1636$-$53 with different models, including
both symmetric and assymmetric line profiles.
They found that, in four observations the primary source of the hard X-rays that reflect off the disc
was the NS surface/boundary layer, and in two observations the primary source was the corona.

Additionally, the Fe line profile in 4U~1636$-$53 shows a blue wing extending to high energies
\citep{Pandel2008}, which indicates a high inclination, even though neither eclipses or dips have been observed in this source.
\cite{Sanna2013} also reported a high inclination of the source in most cases. 
\cite{Sanna2013} tried both phenomenological and reflection models, but none of these models helped solving this high-inclination issue. 

The {\it Nuclear Spectroscopic Telescope Array} ({\Nu}, \citealt{Harrison2013}) is the first focusing high energy (3$-$79 keV)
X-ray observatory. Compared with {\it XMM-Newton}, {\Nu} can simultaneously observe the broad emission line and the Compton 
hump without pileup effects, so it offers an ideal opportunity to study the reflection spectra not only in LMXBs
but also in AGN. 
The good energy resolution and sensitivity of {\Nu} allow
us to better constrain the hard X-ray continuum, identifying the presence of Comptonization and of a cut-off
in the high energy emission. 
Recently, \cite{Ludlam2017} analysed one {\Nu} observation of 4U~1636$-$53 in the hard state and they 
found a high inclination of 76.5$\degree$$-$79.9$\degree$ for a spin parameter of 0.0-0.3, which is consistent with the
inclination derived from the other papers above.
Here we report on another three observations of 4U~1636$-$53 taken with {\Nu}, which are subsequent to the observation in \cite{Ludlam2017}, while the source was in different states. We apply different models to investigate the characteristics of the line and compare those characteristics in different states of the source.

\section{OBSERVATIONS AND DATA REDUCTION}

\begin{figure*}
\centering
\includegraphics[width=8cm,angle=270]{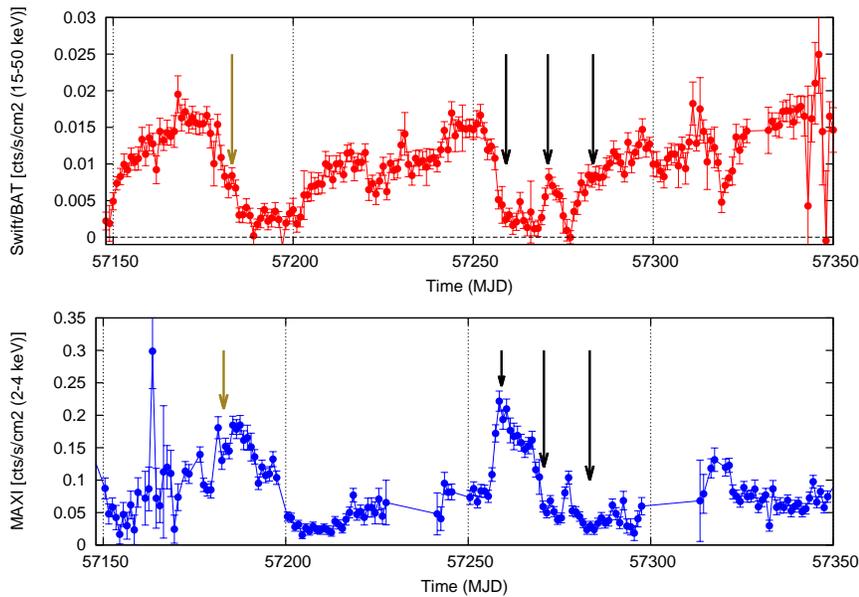}
% \hspace{-2.2em}
\caption{\label{fig:lc}Hard and soft long-term light curves of 4U~1636$-$53. Top and bottom panels show, respectively, the Swift/BAT
  (15$-$50~keV) and the MAXI (2$-$4~keV) light curve of this source. The olive and black arrows mark, respectively, the times
  of the {\Nu} observation used in \protect\cite{Ludlam2017} and the three {\Nu} observations discussed in this paper.}
\end{figure*}

The X-ray data used here consist of three observations with {\Nu} taken between August 25 and September 18 2015.
We report the details of the observations in Table~\ref{tab:1}. 
We marked the time of the three {\Nu} observations presented here in Fig.~\ref{fig:lc}, 
which shows the publicly available
Swift/BAT daily-averaged light curve (15$-$50~keV, top
panel)\footnote{http://swift.gsfc.nasa.gov/results/transients/4U1636-536/}
and the MAXI daily-averaged light curve (2$-$4~keV, lower
panel)\footnote{http://maxi.riken.jp/top/}.
The light curve of 4U~1636$-$53 in Fig.~\ref{fig:lc} shows a $\sim$35$-$40~days long-term evolution \citep{Shih2005,Belloni2007}
related to spectral changes as the system moves between the hard and soft spectral states, 
which indicates that the source evolves from the soft, to the transitional, and the hard state from Obs.~1 to Obs.~3.
In Fig.~\ref{fig:lc} we also marked the observation analysed by \cite{Ludlam2017}, in which, according to these authors,
the source was in the hard state.

We processed the {\Nu} data using the {\Nu} Data Analysis Software (NuSTARDAS) version 1.5.1. \citep{Harrison2013}. 	
We extracted light curves and spectra with the command {\it nuproducts} using a circular extraction region of 100\arcsec~
for both focal plane modules A and B (FPMA/B). We used another similar sized region away from the source, avoiding the stray 
light from a nearby source, as the background spectra.
There were seven type I X-ray bursts in total in the three observations. 
A more detailed discussion of the bursts will be presented in a separated paper.
We created good time intervals (GTIs)  
to eliminate the bursts from the spectra of the persistent emission.
Finally we grouped the spectra with a minimum of 25 counts per spectral bin using the task {\it grppha} within ftools.

\begin{table*}
\caption{\label{tab:1}{\Nu} Observations of 4U 1636$-$53 used in this paper}
\renewcommand{\arraystretch}{1.3}
\footnotesize
\centering
\begin{tabular}{lcccc}
\hline \hline
\multirow{2}{*}{Obervation}&\multirow{2}{*}{Identification Number}&Observation Times (UTC)&\multirow{2}{*}{Exposure (ks)}\\
                                            &                                     &(day.month.year hr:min)&   \\

\hline
Obs.~1& 30102014002&25.08.2015 02:51 - 25.08.2015 18:36 & 27.4$^A$ (27.3$^*$)/27.7$^B$ (27.5$^*$) \\

Obs.~2 & 30102014004&05.09.2015 17:41 - 06.09.2015 11:01 & 30.3$^A$ (30.2$^*$)/30.4$^B$ (30.3$^*$)\\

Obs.~3 & 30102014006&18.09.2015 07:06 - 18.09.2015 23:26 & 28.9$^A$ (28.8$^*$)/29.0$^B$ (28.9$^*$)\\
\hline

\end{tabular}
\begin{flushleft}
$^A$Total exposure time of FPMA of {\Nu};\\
$^B$Total exposure time of FPMB of {\Nu};\\
$^*$Final exposure time excluding X-ray bursts. 
\end{flushleft}
\end{table*}

\section{SPECTRAL ANALYSIS AND RESULTS}

We used the spectral analysis package XSPEC version 12.9.0 to fit the {\Nu} spectra 
of 4U~1636$-$53 between 3 and 79~keV, except for the observation 30102014002.
The spectrum of observation 30102014002 is very soft and is background-dominated at $E>$30~keV, 
resulting in the high-energy data becoming very noisy.
We therefore restricted the spectral fits of observation 30102014002 to the energy range of 3$-$30~keV.
All errors are quoted at the 1~$\sigma$ confidence level unless otherwise specified.

We considered each observation observed by two detectors FPMA and FPMB simultaneously as a group and jointly fitted 
all the groups.
In order to account for flux calibration uncertainties, we included a multiplicative constant in our model.
In all groups we fixed the constant to 1 for FPMA and left it free for FPMB.
We included a {\sc phabs} component in our model to account for the interstellar
absorption along the line of sight to this source. 
When leaving the parameter $N_{\rm H}$ of this component free, it becomes significantly smaller than previously found in this source.
Previous studies of 4U~1636$-$53 with {\it XMM-Newton}, which extend down to $\sim0.5$~keV, have found that $N_{\rm H}$
is about $3~\times~10^{21}~cm^{-2}$ \citep[e.g.][]{Pandel2008}.
{\Nu} data only extend down to 3~keV, and hence we cannot constrain $N_{\rm H}$ from our fits.
We therefore fixed the value of $N_{\rm H} = 3.1 \times 10^{21}~{\rm cm}^{-2}$ \citep{Zhang2017}.

Following previous studies of the continuum spectra of 4U~1636$-$53 \citep[e.g.][]{Ng2010}, we initially used 
a multi-colour disc blackbody component to account for emission from the ion disc \citep[{\sc diskbb} in XSPEC,][]{Mitsuda1984}, 
a single temperature blackbody that represents the emission 
from the NS surface and the boundry layer ({\sc bbody} in XSPEC), and
a thermal comptonisation component \citep[{\sc nthcomp} in XSPEC,][]{Zdziarski1996,Zycki1999}.
Compared to an exponentially cut-off power-law, the {\sc nthcomp} component offers a sharper
high-energy cut-off and a more accurate low-erengy rollover with similar parameters. 
In previous works, using {\it XMM-Newton}, the temperature of the {\sc diskbb} component was {\it $kT_{\rm dbb}~\sim$}
0.2$-$0.8~keV \citep[e.g.][]{Sanna2013, Lyu2014}. Given that {\Nu} only covers the spectrum above 3~keV, it is not possible to constrain
this component with these data. All our fits give equally good results if we exclude the {\sc diskbb}
component from the model. Therefore, we did not include this component in the rest of the analysis. We note that this does not
mean that there is no disc emission in this source (for instance, as we discuss below, the most likely source
of the seed photons of the {\sc nthcomp} component is actually the disc); it is only that {\Nu} data do not allow us
to constrain the direct emission of the disc in this source.

\begin{figure}
\centering
\includegraphics[width=8cm,angle=270]{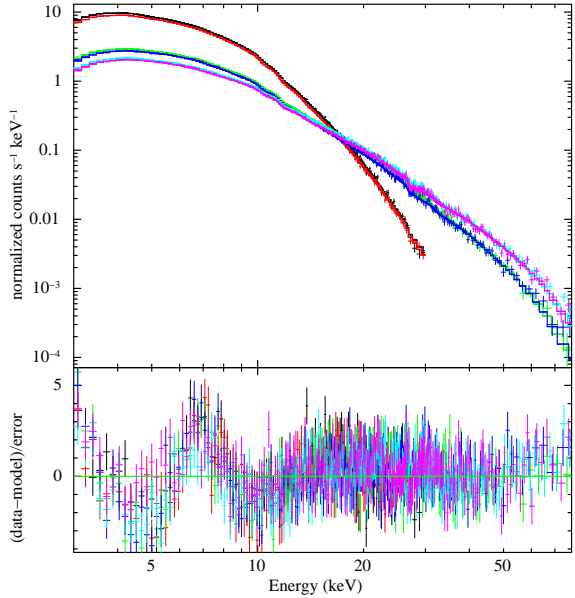}
% \hspace{-2.2em}
\caption{\label{fig:bb+nth}{\Nu} spectra and models for the fit with {\sc const*phabs*(bbody+nthcomp)} for 4U~1636$-$53.
The black, red, green, blue, light blue and magenta lines in the top panel represent each spectrum (FPMA/B) listed from top to bottom in Table~\ref{tab:1}, separately.
The bottom panel shows the residuals in terms of sigmas; the colors are the same as in the top panel. 
The spectra have been rebinned for plotting purposes.}
\end{figure}

The seed photons in the {\sc nthcomp} component could either come from the {\sc diskbb} component 
or the {\sc bbody} component. \cite{Sanna2013} explored the 
origin of the seed photons by linking the seed photon temperature ($kT_{\rm seed}$) in the {\sc nthcomp} component
to either the temperature of the {\sc diskbb} component, $kT_{\rm dbb}$, or to that of the {\sc bbody} component, $kT_{\rm bb}$, respectively,
and they concluded that the seed photons must come from the disc. 
Given that we have no {\sc diskbb} component in our model, we initially set $kT_{\rm seed}$ equal to $kT_{\rm bb}$. 
The {\sc bbody} component in this case became insignificant,
similar to what \cite{Sanna2013} found. We therefore left the $kT_{\rm seed}$
in the {\sc nthcomp} component free to vary with a lower limit at 0.01~keV. 
Following \cite{Sunyaev1980}, the scattering optical depth of the Comptonizing medium, $\tau$, can be calculated 
from the temperature of the Comptonizing electrons, $kT_{e}$, and the power-law photon index, $\Gamma$, as:

\begin{equation}
  \tau = \sqrt{2.25+ \frac{3}{(kT_{\rm e}/511~\rm keV)[( \Gamma +0.5)^2-2.25]} }-1.5.
\end{equation}
After fiting the data with a model containing these components, 
we still found large positive residuals around 5$-$10~keV (see Fig.~\ref{fig:bb+nth}),
which suggests a possible emission line from Fe-K here.
In order to check whether these residuals were due to the continuum model we used,
we replaced the {\sc nthcomp} component by a simple power-law component with a  
high-energy exponential rolloff ({\sc cutoffpl} in XSPEC). We got similar positive residuals in this case as well.

\begin{figure}
\centering
\includegraphics[width=8cm,angle=270]{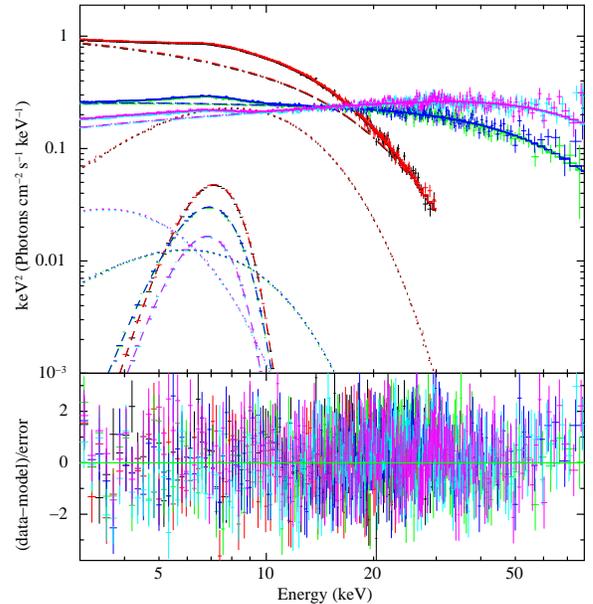}
% \hspace{-2.2em}
\caption{\label{fig:gau}{\Nu} unfolded spectra and models fitted with the model {\sc const*phabs*(bbody+gaussian+nthcomp)} for 4U~1636$-$53.
The colours are the same as in Fig.~\ref{fig:bb+nth}.
The dot, dash and dash-dot lines represent the {\sc bbody}, {\sc gaussian} and {\sc nthcompt} 
components in our model, respectively. 
Notice that in the top panel of the following spectra figures, the $y$-axis always shows $E^{2}f(E)$.
The bottom panel shows the residuals in terms of sigmas. 
The best-fitting parameters are given in Table~\ref{tab:2}.}
\end{figure}

To try and fit these residuals,
we first added a simple Gaussian component to the model.
We constrained the energy of the {\sc gaussian} component to be
between 6.4 and 7~keV (but see below), and left the width ($\sigma$) and normalisation 
($k_{\rm gau}$) free. The entire model we used was {\sc const*phabs*(bbody+gaussian+nthcomp)}, hereafter, Model~1. 
For every component, we linked all the free parameters within each observation.
The best fitting parameters of Model~1 are listed in Table~\ref{tab:2}; the corresponding spectra, individual components and residuals are shown in Fig.~\ref{fig:gau}.

In all three observations the temperature of the seed photons, $kT_{\rm seed}$, in the {\sc nthcomp}
component is not well constrained and is consistent with zero.
The  power-law photon index, $\Gamma_{\rm nth}$, of the {\sc nthcomp} component decreases while 
the cut-off energy, $kT_{\rm e}$, increases from Obs.~1 to Obs.~3.
The optical depth, $\tau$, drops abruptly from Obs.~1 to Obs.~2 and then remains more or less constant 
from Obs.~2 to Obs.~3.
Based on previous spectral analyses of 4U~1636$-$53 \citep[e.g.][]{Sanna2013},
the trend of these parameters implies that the Obs.~1, 2, and 3 
sampled the source, respectively, in the soft, the transitional, and the hard state.
The energy of the {\sc gaussian} component, $E_{\rm gau}$, decreases from $\sim6.7~$keV in Obs.~1 to $\sim$6.4~keV in
Obs.~2 and 3, which means that the disc becomes less ionized.
If we allowed the energy of the line to be below 6.4~keV (in our case we constrained it to be between 5 and 7~keV) 
because of a possible gravitational redshift, we found that the energy of the line in Obs.~1 
does not change significantly, in this case being $6.72\pm0.1$~keV with $\sigma=1.23\pm0.20$,
but the energy of the line in Obs.~2 and 3 decreases to $6.17\pm0.13$~keV with $\sigma=1.56\pm0.12$ and $5.35_{-0.43}^{+0.55}$~keV with $\sigma=1.75\pm0.23$, respectively.
The $kT_{\rm bb}$ goes down with time. The flux of the {\sc gaussian} component, $F_{\rm gau}$,
decreases all the time, whereas the fluxes of the {\sc bbody} and of the {\sc nthcomp} components decrease at the beginning 
and slightly increase in the last observation.
It is apparent that the emission in the 3$-$79~keV range is always dominated by the {\sc nthcomp} component.
Although the fits with a Gaussian line are statistically acceptable, given the broad profile of the {\sc gaussian} ($\sigma$ between 1.2 and 1.5~keV),
we also modeled our data with reflection models that include relativistic effects that affect the profile of the line.

\begin{table}
\caption{\label{tab:2}Best-fitting parameters of the {\Nu} spectra of 4U~1636$-$53 with Model~1}
\renewcommand{\arraystretch}{1.5}
\scriptsize
\centering
\begin{tabular}{lcccccccccccc}
\hline \hline
\multicolumn{2}{c}{\multirow{2}{*}{Component}} & \multicolumn{3}{c}{Model~1}\\
                                &          &Obs.~1&Obs.~2&Obs.~3        \\
\hline
\multicolumn{2}{c}{const}&  \multicolumn{3}{c}{1.00$^f$/1.02$\pm0.01$}\\
% \multicolumn{2}{c}{$N\rm{_{H}}$ ($\rm{10^{22}~cm^{-2}}$)}&\multicolumn{6}{c}{0.31$^f$}\\
\hline

\multirow{3}{*}{{\sc bb}}&$kT_{\rm bb}$ (keV) &2.03$\pm0.02$&1.51$\pm0.07$&0.88$\pm0.02$ \\
&$k_{\rm bb}$~($10^{-3}$) & 5.9$\pm0.2$ &0.3$\pm0.1$&0.8$\pm0.1$                       \\
&\multirow{2}{*}{$F_{\rm bb}$~($10^{-11}$)}&45.5$\pm0.8$&2.3$\pm0.3$&3.2$\pm0.1$\\[-5pt]
&                                          &44.5$\pm1.0$&2.1$\pm0.3$&3.6$\pm0.1$\\
\hline
\multirow{7}{*}{{\sc gaussian}}&$E_{\rm gau}$~(keV) &6.72$\pm0.06$&6.40$_{-0p}^{+0.03}$&6.40$_{-0p}^{+0.03}$\\
&$\sigma$~(keV)&1.23$\pm0.10$&1.35$\pm0.05$&1.22$\pm0.06$\\
&$k_{\rm gau}~(10^{-3})$&3.0$\pm0.3$&2.3$\pm0.2$&1.1$\pm0.1$  \\
&\multirow{2}{*}{$EW~$(keV)}&0.17$\pm0.03$&0.33$_{-0.04}^{+0.10}$&0.27$\pm0.09$\\[-3pt]
&                          &0.17$\pm0.03$&0.71$_{-0.10}^{+0.05}$&0.24$\pm0.07$\\
&\multirow{2}{*}{$F_{\rm gau}$~($10^{-11}$)}&3.25$\pm0.23$&2.26$\pm0.09$&1.02$\pm0.10$ \\[-3pt]
&                                           &3.30$\pm0.25$&2.40$\pm0.09$&1.38$\pm0.06$ \\
\hline
\multirow{5}{*}{{\sc nthcomp}}&$\Gamma_{\rm nth}$ & 2.33$\pm0.02$&2.04$\pm0.01$&1.79$\pm0.01$ \\
&$kT_{\rm e}~(\rm keV)$ & 3.7$\pm0.04$&16.9$\pm0.5$ &20.9$\pm0.7$ \\
&$\tau$ & 7.5$\pm0.3$&3.4$\pm0.1$ &3.7$\pm0.2$\\
&$kT_{\rm seed}~(\rm keV)$ &0.13$_{-0.13}^{+0.10}$ &0.25$_{-0.24}^{+0.05}$ &0.06$_{-0.06}^{+0.32}$\\
&$k_{\rm nth}$ & 1.36$\pm0.08$ & 0.25$\pm0.03$&0.12$\pm0.06$\\ 
&\multirow{2}{*}{$F_{\rm nth}$~($10^{-9}$)}&1.64$\pm0.01$&1.07$\pm0.01$&1.12$\pm0.01$\\[-5pt]
&                                          &1.63$\pm0.01$&1.07$\pm0.01$&1.13$\pm0.01$\\
\hline
&\multirow{2}{*}{Total Flux~$F_{\rm ttl}$~($10^{-9}$)}&2.12$\pm0.01$&1.12$\pm0.01$&1.16$\pm0.01$\\
&                                                     &2.12$\pm0.01$&1.12$\pm0.01$&1.18$\pm0.01$\\
\hline
\multicolumn{2}{c}{$\chi_v^2$ (dof)} &\multicolumn{3}{c}{1.04(4637)}\\
\hline
\end{tabular}
\begin{flushleft}
{\bf Notes.} $k_{\rm bb}$, $k_{\rm gau}$ and $k_{\rm nth}$ are the normalisation of each component 
in units of $\rm photons~keV^{-1}cm^{-2}s^{-1}$.
All the flux as $F_{\rm bb}$, $F_{\rm gau}$, $F_{\rm nth}$ and $F_{\rm ttl}$, represent the unabsorbed flux 
in units of $\rm erg~cm^{-2}s^{-1}$ in the 3$-$79~keV range. 
The symbol, ${p}$, indicates that the energy of the {\sc gaussian} component pegged at the lower limit.
Errors are quoted at 1$\sigma$ confidence level.
\end{flushleft}
\end{table}

\subsection{PHENOMENOLOGICAL REFLECTION MODEL OF THE LINE}
\begin{figure}
\centering
\includegraphics[width=8cm,angle=270]{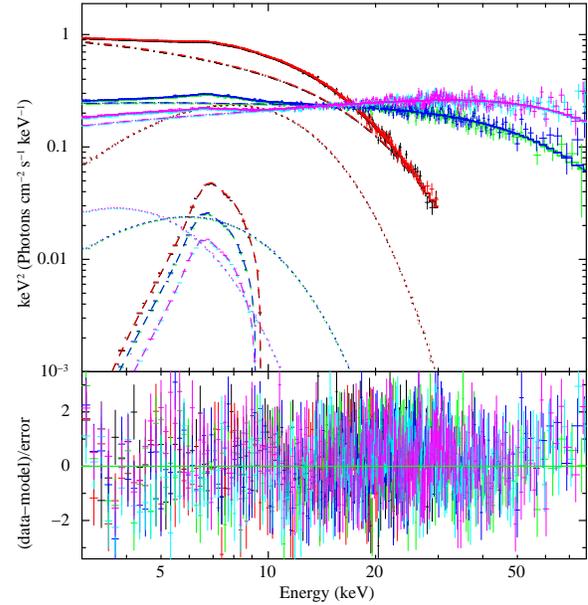}
% \hspace{-2.2em}
\caption{\label{fig:kyr}{\Nu} unfolded spectra and models fitted with the model {\sc const*phabs*(bbody+kyrline+nthcomp)} for 4U~1636$-$53.
Colours and lines are the same as in Fig.~\ref{fig:bb+nth},
except for the dash-dot line representing the {\sc kyrline} component in this model. 
The bottom panel shows the residuals in terms of sigmas. 
The corresponding parameters are given in Table~\ref{tab:3}.}
\end{figure}

Compared with the other popular relativistic iron line models (e.g. {\sc diskline, loar}), {\sc kyrline} \citep{Doviak2004}
allows to set the spin parameter, $a$, to values different from 0 and 1,
and takes the effect of limb darkening in the disc into account.
The fit parameters of the model {\sc kyrline} are the dimensionless spin of the NS, $a$, the inclination of the disc, $i_{\rm kyr}$,
the inner and outer radii of the disc, $R_{in}$ and $R_{out}$, respectively, the rest energy of the line, $E_{\rm kyr}$,
the inner and outer emissivity index, $\alpha$ and $\beta$, respectively, and the normalisation of the line, $k_{\rm kyr}$.      
Assuming that the line is due to iron, from neutral to highly ionized,
we constrained $E_{\rm kry}$ to be between 6.4 and 7~keV.
Following \cite{Braje2000}, the spin parameter is $a=0.47/P$~(ms), which,
for a NS spin of 581~Hz \citep{Zhang1997,Strohmayer2002}, 
gives $a=0.27$, and the smallest possible value for the inner radius of the disc is 
$R_{in}=5.12~R_g$, where $R_{g}=GM/c^2$ \citep{Miller1998}.
We fixed $R_{out}$ to 400~$R_{g}$.
We tied $\alpha$ and $\beta$ to get a single emissivity index. 
We also included an {\sc nthcomp} component
to fit the hard emission of our spectra.
Hereafter, we call Model~2 to the model {\sc const*phabs*(bbody+kyrline+nthcomp)}.
The best-fitting parameters of Model~2 are given in Table~\ref{tab:3}; the corresponding spectra, individual components
and residuals are shown in Fig.~\ref{fig:kyr}.

\begin{figure}
\centering
\includegraphics[width=8cm]{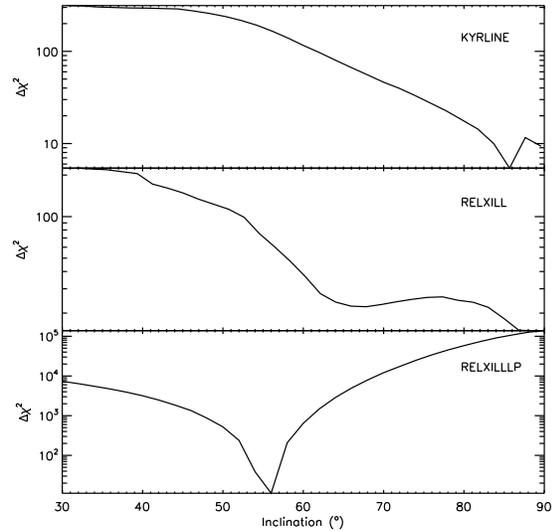}
% \hspace{-2.2em}
\caption{\label{fig:delta_chi} The change in the goodness-of-fit, $\Delta \chi^2$, versus the inclination of different models for
the {\Nu} observations of 4U~1636$-$53. The $\Delta \chi^2$ was calculated using the command {\it steppar} in XSPEC
over 30 steps in the inclination angle.
The $y$-axis is in logarithmic scale. The panels from top to bottom correspond to the best-fitting of Model~1 to 3, respectively.}
\end{figure}

Comparing Tables~\ref{tab:2} and \ref{tab:3}, we see that all the parameters of the continuum
components are more or less the same when we fit the line with either {\sc gaussian} or {\sc kyrline}. 
Only the flux and the equivalent width (EW) of {\sc kyrline} are smaller than those of {\sc gaussian}.
Even though there are less degrees of freedom in Model~2 than in Model~1, the fit does not improve significantly using
{\sc kyrline} compared to {\sc gaussian}. 
Also, as in the case of Model~2 in all observations, $kT_{\rm seed}$ of the {\sc nthcomp}
component is not well constrained and is consistent with zero. 
In all observations the inner disc radius pegs at the lower limit of the model, 
and the emissivity index remains more or less constant in all three observations. 
Remarkably, the inclination is quite high, larger than 80\degree. 
The top panel in Fig.~\ref{fig:delta_chi} shows the $\Delta \chi^2$ of the fit versus the inclination for Model~2.

\begin{figure}
\centering
\includegraphics[width=8cm]{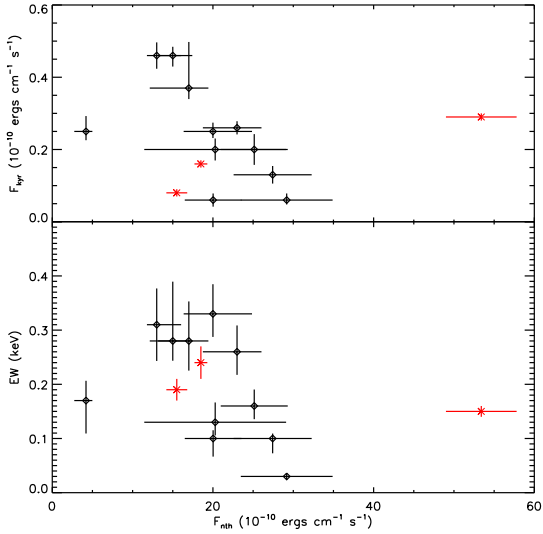}
% \hspace{-2.2em}
\caption{\label{fig:eqw_flux}The flux (top panel) and equivalent width (bottom panel) of the {\sc kyrline} component 
versus the flux (0.5$-$130~keV) of the {\sc nthcomp} component in Model~2 of 4U~1636$-$53.
The three red crosses show the data obtained from {\Nu} observations in this work;
the 11 back diamonds represent the {\it Suzaku} and {\it XMM-Newton/RXTE} data that given by \protect\cite{Lyu2014}. 
Error bars correspond to the 1$\sigma$ uncertainty.}
\end{figure}

In addition, the flux or the EW of the {\sc kyrline} component and the flux of
the {\sc nthcomp} in Model~2 are anti-correlated. 
\cite{Lyu2014} found a similar result in their work, in which they used five observations
from {\it Suzaku} and six observations from {\it XMM-Newton/RXTE}. 
The best-fitting parameters of the {\sc bbody} and {\sc nthcomp} components in Model~1 and 2 here are
consistent with the best-fitting parameters of those same components in \cite{Lyu2014}
when they fitted similar models to {\it XMM-Newton} and {\it RXTE} data of this source.
In Fig.~\ref{fig:eqw_flux}, we plot the flux and EW of the {\sc kyrline}
component vs. the flux of the {\sc nthcomp} component.

\begin{table}
\caption{\label{tab:3}Best-fitting parameters of the {\Nu} spectra of 4U~1636$-$53 with Model~2}
\renewcommand{\arraystretch}{1.5}
\scriptsize
\centering

\begin{tabular}{lcccccccccccc}
\hline \hline
\multicolumn{2}{c}{\multirow{2}{*}{Component}} & \multicolumn{3}{c}{Model~2}\\
                                &          &Obs.~1&Obs.~2&Obs.~3       \\
\hline
\multicolumn{2}{c}{const}&  \multicolumn{3}{c}{1.00$^f$/1.02$\pm0.01$}\\
\hline

\multirow{3}{*}{{\sc bb}}&$kT_{\rm bb}$ (keV)&2.01$\pm0.01$ &1.50$\pm0.04$ &0.94$\pm0.03$ \\
&$k_{\rm bb}$~($10^{-3}$)             &6.1$\pm0.2$&0.6$\pm0.1$&0.8$\pm0.1$ \\
&\multirow{2}{*}{$F_{\rm bb}$~($10^{-11}$)}&46.8$\pm0.6$&4.3$\pm0.2$&3.3$\pm0.1$\\[-5pt]
&                                         &45.9$\pm0.6$&4.1$\pm0.2$&3.8$\pm0.1$\\
\hline
\multirow{8}{*}{{\sc kyrline}}&$i_{\rm kyr}$~($\degree$)&\multicolumn{3}{c}{87.6$\pm0.8$}\\
&$R_{\rm in}/R_{\rm g}$&5.12$_{-0p}^{+0.15}$&5.12$_{-0p}^{+0.10}$&5.12$_{-0p}^{+0.08}$\\
&$\alpha=\beta$&2.36$\pm0.11$&2.43$_{-0.09}^{+0.14}$&2.35$_{-0.09}^{+0.17}$\\
&$E_{\rm kyr}$ (keV) &6.61$\pm0.06$&6.41$_{-0p}^{+0.07}$&6.40$_{-0p}^{+0.04}$\\
&$k_{\rm kyr}~(10^{-3})$&2.73$\pm0.13$&1.55$\pm0.13$&0.89$\pm0.19$\\
&\multirow{2}{*}{$EW_{\rm kyr}~$(keV)}&0.15$\pm0.01$&0.24$\pm0.03$&0.18$\pm0.03$  \\[-3pt]
&                                    &0.15$\pm0.01$&0.25$\pm0.03$&0.21$\pm0.01$ \\
&\multirow{2}{*}{$F_{\rm kyr}$~($10^{-11}$)}&2.91$\pm0.11$&1.55$\pm0.05$&0.76$\pm0.05$\\[-5pt]
&                                           &2.93$\pm0.11$&1.67$\pm0.06$&1.08$\pm0.05$\\
\hline
\multirow{5}{*}{{\sc nthcomp}}&$\Gamma_{\rm nth}$ &2.33$\pm0.02$&2.03$\pm0.01$&1.79$\pm0.01$\\
&$kT_{\rm e}~(\rm keV)$ &3.8$\pm0.1$&16.5$\pm0.5$&21.1$\pm0.8$\\
&$\tau$ & 7.0$\pm0.2$&3.5$\pm0.2$ &3.6$\pm0.2$\\
&$kT_{\rm seed}~(\rm keV)$ &0.14$_{-0.14}^{+0.15}$&0.14$_{-0.14}^{+0.17}$&0.15$_{-0.15}^{+0.20}$\\
&$k_{\rm nth}$ &1.35$_{-0.18}^{+0.04}$&0.26$\pm0.01$&0.12$_{-0.01}^{+0.01}$\\ 
&\multirow{2}{*}{$F_{\rm nth}$~($10^{-9}$)}&1.63$\pm0.01$&1.06$\pm0.01$&1.12$\pm0.01$\\[-5pt]
&                                          &1.62$\pm0.01$&1.06$\pm0.01$&1.13$\pm0.01$\\ 
\hline
&\multirow{2}{*}{Total Flux~$F_{\rm ttl}$~($10^{-9}$)}&2.12$\pm0.01$&1.12$\pm0.01$&1.16$\pm0.01$\\
&                                                     &2.12$\pm0.01$&1.12$\pm0.01$&1.18$\pm0.01$\\
\hline
\multicolumn{2}{c}{$\chi_v^2$ (dof)} &\multicolumn{3}{c}{1.04(4633)}\\
\hline
\end{tabular}
\begin{flushleft}
{\bf Notes.} Units are the same as in Table~\ref{tab:2}. Errors are quoted at 1$\sigma$ confidence level. 
The inclination of the {\sc kyrline} component is linked across the three observations.
\end{flushleft}
\end{table}

\subsection{FULL REFLECTION MODELS}

\begin{figure}
\centering
\includegraphics[width=8cm,angle=270]{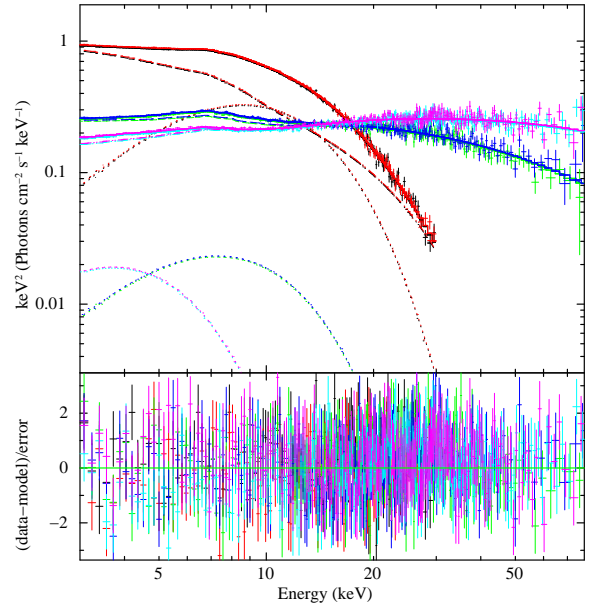}
% \hspace{-2.2em}
\caption{\label{fig:rel}{\Nu} unfolded spectra and models fitted with the model {\sc const*phabs*(bbody+relxill)} for 4U~1636$-$53.
Colours are the same in Fig.~\ref{fig:bb+nth},
The dot and dash-dot line represent the {\sc bbody} and {\sc relxill} components in this model. 
The bottom panel shows the residuals in terms of sigmas. 
The corresponding parameters are given in Table~\ref{tab:4}.}
\end{figure}

Even though there is no clear reflection hump at high energies (above 10~keV) in
the residuals of Fig.~\ref{fig:bb+nth}, 
the presence of a broad iron line suggests that reflection off the ion disc is important.
We therefore fitted the self-consistent relativistic
reflection model {\sc relxill} \citep{Dauser2013,Dauser2016A}, which describes not only the reflection part, but also a direct power-law component. 
However, the power-law continuum within {\sc relxill} differs from that in {\sc nthcomp}. 
In particular, the high-energy cut-off in these two models behaves differently \citep{Garc2015}.
The parameters in this model are the inclination of the 
disc, $i_{\rm rel}$, the inner and outer radii of the disc, $R_{in}$ and $R_{out}$, respectively,
the inner and outer emissivity indexes, $q_{in}$ and $q_{out}$, respectively,
the break radius, $R_{break}$, between the two emissivity laws, the spin parameter, $a$, the redshift to the source, $z$,
which we fixed to 0,
the photon index of the power-law, $\Gamma$, the cut-off energy of the power-law, $E_{cut}$,
the ionization parameter, $\xi$,
the iron abundance, $A_{Fe}$, which we fixed to solar, the reflection fraction, $f_{refl}$, and the normalisation, $k_{\rm rel}$.
The overall model becomes {\sc const*phabs*(bbody+relxill)}, hereafter Model~3.
As for {\sc kyrline}, we fixed $a$ to 0.27, and hence $R_{in}$ was forced to be larger than 5.12~$R_g$.
$R_{out}$ was fixed at 400~$R_g$, and $q_{in}$ and $q_{out}$ were linked to vary together.
The best-fitting parameters of Model~3 are given in Table~\ref{tab:4}; the corresponding spectra, individual components 
and residuals are shown in Fig.~\ref{fig:rel}.

\begin{figure}
\centering
\includegraphics[width=8cm,angle=270]{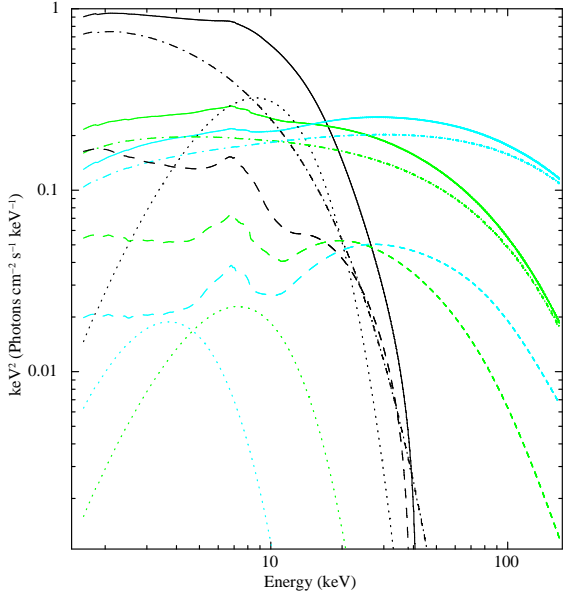}
% \hspace{-2.2em}
\caption{\label{fig:eemo}The unfolded best-fitting model {\sc const*phabs*(bbody+relxill)} to the spectra of 4U~1636$-$53. 
The black, green and light blue lines correspond to each spectrum (only FPMA) listed from top to bottom in Table~\ref{tab:1}.
The dot lines represent the {\sc bbody} component. The dash and dot-dash lines represent the reflection 
and power-law spectra within the {\sc relxill} model, respectively.
 }
\end{figure}

Most of the parameters of Model~3 follow the same trend as those of the other models.
The inclination, $i_{\rm rel}$, in Model~3 is still extremely high, consistent with $i_{\rm kyr}$ in Model~2.
The inner radius of the disc, $R_{\rm in}$, and cut-off energy, $E_{\rm cut}$, of each observation in Model~3, 
however, are larger than the corresponding ones in Model~2.
The optical depth $\tau$ drops abruptly from Obs.~1 to Obs.~2, then stays almost constant,
but the values of the optical depth in Model~3 are smaller than those in Model~2.
The high reflection fraction, $f_{refl}$, indicates that reflection features in these spectra are important.
We, therefore, plot the reflection and power-law components of Model~3 separately in Fig.~\ref{fig:eemo}.
In order to show the reflected part clearly, 
we only plot the three unfolded model spectra of FPMA in Fig.~\ref{fig:eemo}.
In the middle panel of Fig.~\ref{fig:delta_chi} we show the $\Delta \chi^2$ of the fit versus the inclination for Model~3.

\begin{table}
\caption{\label{tab:4}Best-fitting parameters of the {\Nu} spectra of 4U~1636$-$53 with Model~3}
\renewcommand{\arraystretch}{1.5}
\scriptsize
\centering
\begin{tabular}{lcccccccccccc}
\hline \hline
\multicolumn{2}{c}{\multirow{2}{*}{Component}} & \multicolumn{3}{c}{Model~3}\\
                                &          &Obs.~1&Obs.~2&Obs.~3         \\
\hline
\multicolumn{2}{c}{const}&  \multicolumn{3}{c}{1.00$^f$/1.02$\pm0.01$}\\
% \multicolumn{2}{c}{$N\rm{_{H}}$ ($\rm{10^{22}~cm^{-2}}$)}&\multicolumn{6}{c}{0.31$^f$}\\
\hline
\multirow{3}{*}{{\sc bb}}&$kT_{\rm bb}$ (keV)&2.19$\pm0.02$ &1.85$\pm0.10$ &0.93$\pm0.04$\\
&$k_{\rm bb}$~($10^{-3}$)             &8.4$\pm0.1$&0.6$\pm0.1$&0.5$\pm0.1$\\
&\multirow{2}{*}{$F_{\rm bb}$~($10^{-11}$)}&65.8$\pm0.6$&4.5$\pm0.2$&2.2$\pm0.3$\\[-5pt]
&                                          &64.6$\pm0.5$&4.3$\pm0.3$&2.6$\pm0.3$\\
\hline
\multirow{12}{*}{{\sc relxill}}&$i_{\rm rel}~(\degree)$& \multicolumn{3}{c}{88.0$\pm0.3$}\\
&$R_{\rm in}/R_{\rm g}$&5.8$\pm0.7$&16.1$_{-2.7}^{+4.3}$&16.3$_{-4.5}^{+15.8}$\\
&$q_{\rm in}=q_{\rm out}$ &2.2$\pm0.2$&5.0$_{-1.4}^{+4.9}$&4.0$_{-1.0}^{+2.7}$\\
% &$a$& \multicolumn{3}{c}{0.27$^{f}$}\\
&$\Gamma_{\rm rel}$&2.01$\pm0.05$&1.97$\pm0.02$&1.78$\pm0.04$\\
&$E_{\rm cut}$~(keV)&6.5$_{-1.2}^{+0.4}$&62.8$_{-3.8}^{+4.4}$&136.0$_{-21.7}^{+28.8}$\\
&$\tau$ & 6.3$\pm0.9$&1.4$\pm0.1$ &1.0$\pm0.2$ \\
&$log(\xi)$&3.3$\pm0.3$&3.1$\pm0.1$&2.9$\pm0.1$\\
&$refl\_frac$&0.9$\pm0.1$&1.7$\pm0.1$&1.2$\pm0.1$\\
&$k_{\rm rel}~(10^{-3})$&8.6$\pm0.5$&2.5$\pm0.1$&2.3$\pm0.1$\\
% &\multirow{2}{*}{$F_{\rm pl}$~($10^{-9}$)}&1.1$\pm0.2$&0.8$\pm0.05$&1.0$\pm0.05$\\[-5pt]
% &                                         &1.1$\pm0.2$&0.9$\pm0.05$&1.0$\pm0.06$\\
&\multirow{2}{*}{$F_{\rm rel}$~($10^{-9}$)}&1.5$\pm0.01$&1.1$\pm0.01$&1.1$\pm0.01$\\[-5pt]
&                                          &1.5$\pm0.01$&1.1$\pm0.01$&1.2$\pm0.01$\\
\hline
&\multirow{2}{*}{Total flux~$F_{\rm ttl}$~($10^{-9}$)}&2.12$\pm 0.01$ &1.13$\pm 0.01$&1.17$\pm 0.01$\\[-5pt]
&                                                     &2.11$\pm 0.01$ &1.12$\pm 0.01$&1.19$\pm 0.01$\\
\hline
\multicolumn{2}{c}{$\chi_v^2$ (dof)} &\multicolumn{3}{c}{1.03(4636)}\\

\hline
\end{tabular}
\begin{flushleft}
{\bf Notes.} Units are the same as in Table~\ref{tab:2}. Errors are quoted at 1$\sigma$ confidence level.
The inclination of the {\sc relxill} component is linked across the three observations.
\end{flushleft}
\end{table}

In order to investigate the ion geometry of 4U~1636$-$53, we fitted the spectra with 
another reflection model {\sc relxilllp} \citep{Garc2014,Dauser2016B}, which assumes that the corona is a point source located at a height above the accretion disc along the spin axis of the compact object. 
For these fits, {\sc relxilllp} takes the place of {\sc relxill} so that the model becomes {\sc const*phabs*(bbody+relxilllp)}, hereafter Model~4.
Most of the parameters of {\sc relxilllp} are the same as those in {\sc relxill}
but, instead of the inner and outer emissivity indices, {\sc relxilllp} contains one more parameter, 
$h$, which is the height of the corona.
We set all the common parameters of {\sc relxilllp} to the values that we used in {\sc relxill}:
$a$ to 0.27 and $R_{out}$ to 400~$R_g$.
We successfully modeled the spectra with a reasonable inclination $i_{relp} \sim56\degree$.
The best-fitting parameters of Model~4 are presented in Table~\ref{tab:4}.
Most of the parameters of Model~4 follow the same trend as those of Model~3.
The $R_{in}$ and $\tau$ are smaller in Model~4 than in Model~3, but the normalisation, $k_{relp}$,
is higher than $k_{rel}$. 
The bottom panel of Fig.~\ref{fig:delta_chi} shows the $\Delta \chi^2$ of the fit versus the inclination for Model~4.

\begin{table}
\caption{\label{tab:5}Best-fitting parameters of the {\Nu} spectra of 4U~1636$-$53 with Model~4}
\renewcommand{\arraystretch}{1.5}
\scriptsize
\centering
\begin{tabular}{lcccccccccccc}
\hline \hline
\multicolumn{2}{c}{\multirow{2}{*}{Component}} & \multicolumn{3}{c}{Model~4}\\
                                &          &Obs.~1&Obs.~2&Obs.~3         \\
\hline
\multicolumn{2}{c}{const}&  \multicolumn{3}{c}{1.00$^f$/1.02$\pm0.01$}\\
\hline
\multirow{3}{*}{{\sc bb}}&$kT_{\rm bb}$ (keV)&2.13$\pm0.01$ &1.82$\pm0.02$ &0.93$\pm0.02$\\
&$k_{\rm bb}$~($10^{-3}$)             &9.2$\pm0.02$&0.6$\pm0.01$&0.5$\pm0.01$\\
&\multirow{2}{*}{$F_{\rm bb}$~($10^{-11}$)}&71.3$\pm0.07$&4.4$\pm0.05$&2.0$\pm0.03$\\[-5pt]
&                                          &64.6$\pm0.07$&4.4$\pm0.05$&2.5$\pm0.03$\\
\hline
\multirow{12}{*}{{\sc relxilllp}}&$i_{\rm relp}~(\degree)$& \multicolumn{3}{c}{55.7$\pm0.2$}\\
&$h/R_{\rm g}$&2.3$\pm0.2$&2.5$\pm0.1$&2.8$_{-0.3}^{+0.1}$\\
&$R_{\rm in}/R_{\rm g}$&5.7$\pm0.07$&10.3$\pm0.04$&11.4$\pm0.08$\\
&$\Gamma_{\rm relp}$&2.19$\pm0.01$&1.93$\pm0.01$&1.76$\pm0.01$\\
&$E_{\rm cut}$~(keV)&7.9$\pm0.05$&61.5$\pm0.6$&135.9$\pm0.7$\\
&$\tau$ & 4.9$\pm0.04$&1.5$\pm0.02$ &0.9$\pm0.01$\\
&$log(\xi)$&4.4$\pm0.03$&3.4$\pm0.03$&3.1$\pm0.06$\\
&$k_{\rm relp}~(10^{-3})$&289.9$_{-0.3}^{+332}$&46.7$\pm0.05$&21.3$\pm0.03$\\
% &\multirow{2}{*}{$F_{\rm pl}$~($10^{-9}$)}&1.1$\pm0.2$&0.8$\pm0.05$&1.0$\pm0.05$\\[-5pt]
% &                                         &1.1$\pm0.2$&0.9$\pm0.05$&1.0$\pm0.06$\\
&\multirow{2}{*}{$F_{\rm relp}$~($10^{-9}$)}&1.4$\pm0.01$&1.1$\pm0.01$&1.2$\pm0.01$\\[-5pt]
&                                           &1.4$\pm0.01$&1.1$\pm0.01$&1.2$\pm0.01$\\
\hline
&\multirow{2}{*}{Total flux~$F_{\rm ttl}$~($10^{-9}$)}&2.12$\pm 0.01$ &1.13$\pm 0.01$&1.17$\pm 0.01$\\[-5pt]
&                                                     &2.11$\pm 0.01$ &1.12$\pm 0.01$&1.19$\pm 0.01$\\
\hline
\multicolumn{2}{c}{$\chi_v^2$ (dof)} &\multicolumn{3}{c}{1.04(4636)}\\

\hline
\end{tabular}
\begin{flushleft}
{\bf Notes.} Units are the same as in Table~\ref{tab:2}.
The inclination of the {\sc relxilllp} component is linked across the three observations.
\end{flushleft}
\end{table}

\section{DISCUSSION}
We analysed three {\Nu} observations of the NS LMXB 4U~1636$-$53 in different states
and found prominent positive broad residuals around 5$-$10~keV in all {\Nu} spectra, 
which indicates possible emission reflected off an accretion disc. 
We applied four different models to fit the residuals in the spectra, which are: A simple symmetric model, {\sc gaussian},
a relativistically broadened emission-line model, {\sc kyrline}, and two models including relativistically smeared and ionized reflection off the accretion disc, {\sc relxill} and {\sc relxilllp}. All models fitted the data well,
although {\sc kyrline} and {\sc relxill} yield an inclination of the accretion disc, $\sim88\degree$,
which is at odds with the fact that no dips or eclipses have been observed in this source. 
The {\sc relxilllp} model, however, gives a reasonable inclination of $\sim56\degree$. Additionally, the flux
and the equivalent width of the emission line are anti-correlated with the flux 
of the hard illuminating source in Model~2.

Previous work on modelling the reflection spectrum of 4U~1636$-$53 have found high 
inclination angles \citep[e.g.][]{Pandel2008, Sanna2013}.
By modelling three {\it XMM-Newton} spectra with the
{\sc diskline} component, which describes relativistically broadened line emission
from a disc around a non-rotating black hole,
\cite{Pandel2008} reported that the inclination in all cases is larger than 64$\degree$ and consistent with 90$\degree$.
\cite{Sanna2013} analysed six {\it XMM-Newton} observations 
and found that most of them give high inclination values.
Fitting the {\sc kyrline} model to the {\Nu} data, as in the case of \cite{Sanna2013}, we also
found an inclination of $\sim88\degree$. 
In this case, contrary to the case of the {\sc XMM-Newton}
data, this cannot be due to pileup or similar calibration issues.

We also modeled the data with two relativistically blurred reflection models, {\sc relxill} and {\sc relxilllp}.
Compared with angle-averaged reflection models of the line, {\sc relxill} and {\sc relxilllp} 
calculate the reflection fraction, relativistic blurring and angle-dependent reflection spectrum for
different coronal heights self consistently.
The best-fitting inclination angle in {\sc relxill} is still higher than $80\degree$, similar to that in {\sc kyrline}.
\cite{Ludlam2017} applied the same {\sc relxill} model to one {\Nu} observation taken before 
our observations and they also obtained a high inclination of 76.5$\degree$$-$79.9$\degree$.
The best-fitting inclination angle is reasonable in {\sc relxilllp}, $\sim56\degree$.
{\sc relxilllp} assumes a lamp post geometry of the primary source of the illuminating hard X-rays.
In black hole systems, the reflection fraction in {\sc relxilllp} describes how much flux is emitted towards the disc compared to how much
is emitted directly to the observer. Therefore the fraction of photons hitting 
the accretion disc can be directly measured, making it possible to set constraints on the geometry of the system.
The {\sc relxill} model does not assume any geometry and does
not take any relativistic boosting effects into account \citep{Dauser2016B}.
A further exploration of the reason why {\sc relxilllp} gives a more reasonable inclination angle is beyond the scope of this work.

Using {\sc relxilllp}, we found that the primary source is located close to the NS, at a height of h~$\sim$2$-$3~$R_{g}$,
which is consistent with the fact that  
in similar accreting systems (black holes and AGNs, e.g. \citealt{Dauser2013,Fabian2014})
the corona is compact.
Alternatively, in a NS system, the small height could also
refer to the boundary layer between the
accretion disc and the NS surface as the primary source of the illuminating hard X-rays (see \citealt{Sanna2013}).
Additionally, different from other sources \citep[e.g.][]{Parker2014,Ludlam2016}, 
the iron emission line that dominates the emission at 5$-$10~keV of the reflection spectra of 4U~1636$-$53 
is stronger than the Compton hump that dominates the emission at above 10~keV, 
especially in Obs.~1 (see Fig.~\ref{fig:eemo}).
\cite{Dauser2014} suggested that high spin sources produce strong relativistic reflection features. 
They gave the maximum possible reflection fraction as a function of spin in Fig.~3 of their paper.
Based on the frequency of 4U~1636$-$53, we fixed the spin at 0.27 in this work (see \S 3.1).
As for a spin of 0.27, the corresponding maximum reflection fraction is $\sim$1.2 in \cite{Dauser2014}, which is consistent with  our reflection fraction values in Table~\ref{tab:4}.

In {\sc rexill/lp}, the illuminating source is assumed to be a corona, which is described as a power-law with 
a high-energy cut-off. 
Given this assumption in {\sc rexill/lp}, the corona is responsible for the main contribution of the reflected spectra in Obs.~2 and 3.
As for Obs.~1, 4U~1636$-$53 is likely in the soft state and the corresponding $E_{cut}$ is 
around 7$-$8~keV. The low value of the $E_{cut}$ indicates that the illuminating source that produces the reflection component in Obs.~1 may not be the corona.
\cite{Sanna2013} reported that in two out of six observations (Obs.~2 and 6 in their work) 
the illuminating source is essentially the corona, whereas in the other four observations the main illuminating source is the
surface/boundary layer.
Obs.~2 in their work is also in the soft state and the cut-off energy of the component that represented the corona was $E_{cut}=9.5_{-0.8}^{+0.9}$~keV. 
Therefore, we can not conclude whether the primary source in Obs.~1 is the corona or the NS surface/boundary layer, only based
on the low value of the cut-off energy.

In most cases, the temperature of the {\sc bbody} component is higher than 1~keV in LMXBs \citep[e.g.][]{Cackett2010,Ng2010, Lyu2014}.
However, the $kT_{\rm bb}$ in Obs.3 is always below 1~keV in all of our models.
In order to test whether this is due to the lack of a {\sc diskbb} component,
we added a {\sc diskbb} component in our model, even if it is not required by the data (see \citealt{Sanna2013}).
Given the lack of data below 3~keV, we cannot constrain $kT_{\rm dbb}$.
We therefore assumed an average temperature of 0.5~keV \citep{Sanna2013},
and fixed it in all observations; we set the normalisation free to vary but linked them within each observation.
For instance in Model~1, $kT_{\rm bb}$ increased to $1.92\pm0.10$, $1.50\pm0.03$, $1.06\pm0.08$~keV in Obs.~1, 2 and 3, respectively.
As we suspected, the temperature of the {\sc bbody} component is affected by the presence/absence of a disc component.  
Especially in the hard state, $kT_{\rm dbb}$ can be very low , around $\sim$0.2~keV \citep{Sanna2013},
therefore, the {\sc bbody} component shifts to lower temperatures to compensate for the emission of the accretion disc.
This may be the reason why the $kT_{\rm bb}$ in Obs.~3 is so low. 
Actually, the absence of a {\sc diskbb} component in Model~1, affects not only the {\sc bbody} component, but also the {\sc gaussian} component. As we mentioned in \S 3, if we allowed the energy of the line in the {\sc gaussian} component to be below 6.4~keV, the energy of the line in Obs.~2 and 3 decreases, especially in Obs.~3. Using the command {\it steppar} in XSPEC we found that, when we fit the line with a {\sc gaussian}, the energy of the line is correlated with the $kT_{\rm bb}$ in Obs.~3. On the contrary, there is no correlation between the energy of the line in the {\sc kyrline} component and the $kT_{\rm bb}$ in Model~2. These results indicate that the {\sc gaussian} component in Model~1 is very sensitive to the lack of a {\sc diskbb} component.

\cite{Shih2005} reported a $\sim$40~d period in the {\it RXTE}/ASM light curve of 4U~1636$-$53, which 
they interpret as accretion rate variability due to the X-ray irradiation of the disc.
As the X-ray luminosity decreases, the accretion disc is not fully ionized. As a consequence, the outer regions of the disc cool down and thereby the overall mass accretion decreases, subsequently leading to an X-ray minimum.
The inner edge of the disc recedes as a result of the mass accretion reducing in the inner regions because the high-density disc material there will be exhausted and likely be replaced by a hot corona. 
The three {\Nu} observations analysed here were taken over a few days covering more or less the full $\sim$40~d period.
The evolution of our spectral parameters supports the interpretation of \cite{Shih2005}.
The photon index, $\Gamma$, in all models decreases and the cutt-off energy, $E_{cut}$,
increases from Obs.~1 to Obs.~3, which indicates that the system evolves from the soft, to the transitional,
and finally to the hard state (see \citealt{Sanna2013}).
The {\sc bbody} component weakens dramatically from Obs.~1 to Obs.~3 \citep[e.g.][]{Lyu2014}, which matches the picture above.
In priciple, the parameters of the {\sc bbody} component do not have a clear correlation with the source state.
However, keeping in mind the possiblity that the {\sc bbody} component is partly fitting the emission of
the {\sc diskbb} component, the temperature of the {\sc bbody} component decreases from Obs.~1 to Obs.~3, 
probably due to a drop of the temperature of the {\sc diskbb} component (see above).
Besides that, the reflection continuum also shows a strong correlation with the source state.
According to the standard accretion disc model, as mass accretion rate decreases the disc moves outwards \citep[e.g.][]{Esin1997}.
The inner disc radius, both in {\sc relxill} and {\sc relxilllp}, increases from Obs.~1 to Obs.~3.
As the mass accretion rate decreases, the disc becomes less ionized, resulting in 
the $\xi$ and the energy of the line, $E_{\rm gau}$ and $E_{\rm kyr}$, dropping.

We also found that the flux and the EW of the emission line when fitted with the 
model {\sc kyrline} is anti-correlated with the flux of the {\sc nthcomp} component in Model~2.
\cite{Lyu2014} found that the flux and the EW of the iron line first increase and then decrease as the flux of the
Comptonized component increases when the flux of the Comptonized component is higher than $15~\times10^{-10}\rm erg~cm^{-2}~s^{-1}$.
All the fluxes of the {\sc kyrline} in Model~2 fall into this region of the plot.
\cite{Lyu2014} explained this anti-correlation either by gravitational light bending of the primary source, or 
by changes in the ionization states of the accretion disc.
In the light-bending model \citep{Miniutti2004}, the reflection fraction is correlated to the height of the primary
source above the disk. When the source height is small, within a few $R_{g}$
of the disc, relativistic light bending results in a small
fraction of the emitted photons escaping to infinity and a large fraction of the emitted photons bent
towards the disc. 
The height, $h$, of the corona in Model~4 supports this idea as well.

\section{conclusions}
We modelled the spectra of three {\Nu} observations of the source 4U1636$-$53 in different states.
Four models fitted all spectra equally well but with different line profiles.
Even though the simplest symmetric {\sc gaussian} fitted the data well, the breadth of the line, $\sigma>$1.22~keV,
is unlikely to be produced only by Compton broadening.
Both the phenomenological model {\sc kyrline} and the reflection model {\sc relxill} 
gave an unrealistically high inclination of the accretion disc.
Given that this is the first report on the reflection spectrum of {\Nu} data of 4U1636$-$53,
the high inclination from {\sc kyrline} at least excludes the possible effect of calibration uncertainties of the 
{\it XMM-Newton} data which yielded a similarly high inclination (see \citealt{Sanna2013}).
We find a reasonable inclination from the lamp post reflection model {\sc relxilllp}.
In addition, we provide a possible explanation as to why the temperature of {\sc bbody} is lower than 1~keV in this work.
We also explored the variation of the direct and reflection comtinuum as a function of the
source state. We find and confirm that most of the spectral parameters in 4U~1636$-$53 are strongly correlated with the source state.

\section*{ACKNOWLEDGEMENTS}
We thank the referee for many invaluable comments and suggestions.
We also thank, Thomas Dauser, for the kind explanation of their models, {\sc relxill} and {\sc relxilllp}. 
DA acknowledges support from the Royal Society. TMB acknowledges financial contribution from the agreement ASI-INAF I/037/12/0.
This research is based on
the data from the {\Nu} mission, a project led by the California
Institute of Technology, managed by the Jet Propulsion Laboratory and funded by NASA.
This work has made use of data from the High Energy Astrophysics Science Archive Research Center (HEASARC), provided by NASA/Goddard Space Flight Center (GSFC).

\bibliographystyle{mn2e2}
\bibliography{ref_2015.bib}
\end{document}